\documentclass[aps,prl,reprint,superscriptaddress,longbibliography]{revtex4-1}

\usepackage{graphicx}
\usepackage{dcolumn}
\usepackage{amsmath}
\usepackage{multirow} 
\usepackage{floatrow}
\usepackage{physics} 
\usepackage[normalem]{ulem}
\usepackage{enumitem}

\begin{document}

\title{Entangling Bosonic Modes via an Engineered Exchange Interaction}

\author{Yvonne~Y.~Gao}
\email[E-mail: ]{yvonne.gao@yale.edu}
\thanks{These two authors contributed equally to this work.}
\affiliation{Departments of Physics and Applied Physics, Yale University, New Haven, Connecticut 06520, USA}
\affiliation{Yale Quantum Institute, Yale University, New Haven, Connecticut 06511, USA}
\author{Brian~J.~Lester}
\email[E-mail: ]{brian.lester@yale.edu}
\thanks{These two authors contributed equally to this work.}
\affiliation{Departments of Physics and Applied Physics, Yale University, New Haven, Connecticut 06520, USA}
\affiliation{Yale Quantum Institute, Yale University, New Haven, Connecticut 06511, USA}
\author{Kevin~Chou}
\affiliation{Departments of Physics and Applied Physics, Yale University, New Haven, Connecticut 06520, USA}
\affiliation{Yale Quantum Institute, Yale University, New Haven, Connecticut 06511, USA}
\author{Luigi~Frunzio}
\affiliation{Departments of Physics and Applied Physics, Yale University, New Haven, Connecticut 06520, USA}
\affiliation{Yale Quantum Institute, Yale University, New Haven, Connecticut 06511, USA}
\author{Michel~H.~Devoret}
\affiliation{Departments of Physics and Applied Physics, Yale University, New Haven, Connecticut 06520, USA}
\affiliation{Yale Quantum Institute, Yale University, New Haven, Connecticut 06511, USA}
\author{Liang~Jiang}
\affiliation{Departments of Physics and Applied Physics, Yale University, New Haven, Connecticut 06520, USA}
\affiliation{Yale Quantum Institute, Yale University, New Haven, Connecticut 06511, USA}
\author{S.~M.~Girvin}
\affiliation{Departments of Physics and Applied Physics, Yale University, New Haven, Connecticut 06520, USA}
\affiliation{Yale Quantum Institute, Yale University, New Haven, Connecticut 06511, USA}
\author{Robert~J.~Schoelkopf}
\email[E-mail: ]{robert.schoelkopf@yale.edu}
\affiliation{Departments of Physics and Applied Physics, Yale University, New Haven, Connecticut 06520, USA}
\affiliation{Yale Quantum Institute, Yale University, New Haven, Connecticut 06511, USA}

\begin{abstract}
The realization of robust universal quantum computation with any platform ultimately requires both the coherent storage of quantum information and (at least) one entangling operation between individual elements.  The use of continuous-variable bosonic modes as the quantum element is a promising route to preserve the coherence of quantum information against naturally-occurring errors.  However, operations between bosonic modes can be challenging.  In analogy to the exchange interaction between discrete-variable spin systems, the exponential-SWAP unitary [$\mathbf{U}_{\mathrm{E}}\left(\theta_c\right)$] can coherently transfer the states between two bosonic modes, regardless of the chosen encoding, realizing a deterministic entangling operation for certain $\theta_c$.  Here, we develop an efficient circuit to implement $\mathbf{U}_{\mathrm{E}}\left(\theta_c\right)$ and realize the operation in a three-dimensional circuit QED architecture.  We demonstrate high-quality deterministic entanglement between two cavity modes with several different encodings.  Our results provide a crucial primitive necessary for universal quantum computation using bosonic modes.  
\end{abstract}

\date{\today}

\maketitle

Quantum computation presents an exciting new paradigm for information processing, with promising applications in both fundamental exploration of nature and revolutionary technological advancements. A robust universal quantum computer can be realized with any well-controlled quantum system, but a successful platform will ultimately require the combination of highly-coherent, error-correctable quantum elements and at least one entangling operation between them.  Generally, quantum information can be stored in either discrete- or continuous-variable quantum systems. In the former, quantum information is encoded using two discrete levels of a system, which is the textbook model of quantum computation~\cite{nielsen_quantum_2000}.  Deterministic entangling gates can be implemented using a direct interaction between the elements, but the overhead associated with implementing error-correction protocols in these systems to prolong their quantum coherence is significant~\cite{steane_error_1996,fowler_surface_2012}.  In contrast, continuous-variable quantum systems, e.g., a harmonic oscillator, can take advantage of efficient quantum error correction protocols that encode information in the larger available Hilbert space of each element~\cite{chuang_bosonic_1997,gottesman_encoding_2001, sanders_from-qubits_2002}. However, such encoded states typically have no direct couplings, making deterministic entangling operations between them particularly challenging.

%--------Figure 1:--------------
\begin{figure*}[t!]
	\centering
	\floatbox[{\capbeside\thisfloatsetup{capbesideposition={right,center},capbesidewidth=5.4cm}}]{figure}[\FBwidth]
	{\includegraphics[scale=1]{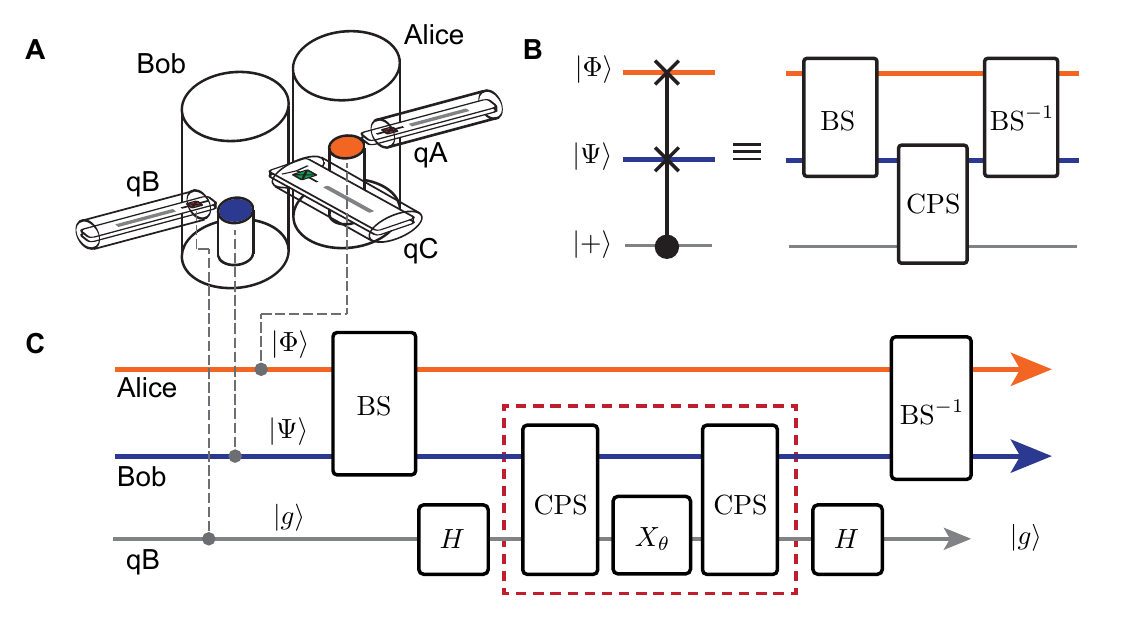}}
	{\caption{\textbf{Sketch (not to scale) of device architecture and experimental protocol}.
		\textbf{(A)} Cartoon of the 3D cQED system used to realize the quantum Fredkin gate and eSWAP operations between two bosonic modes Alice and Bob. 
		\textbf{(B)} Decomposition of the Fredkin gate into two 50:50 beamsplitters and a controlled phase shift (CPS). The CPS is realized using the dispersive coupling between one of the cavity modes and a transmon ancilla as described by the unitary $\mathbf{U}_{\mathrm{CPS}} = |g\rangle\langle g|\otimes \mathbf{I} + |e\rangle\langle e|\otimes e^{i\pi\mathbf{n}}$. 
		\textbf{(C)} Simplified Circuit to realize the eSWAP unitary between two bosonic modes controlled by an ancillary transmon~\cite{supplement}.}
	\label{circuits}}
\end{figure*}
%----------------------------------

Within the circuit quantum electrodynamics (cQED) framework, multiphoton states of superconducting cavities are used to encode continuous-variable quantum information.  In particular, three-dimensional (3D) microwave cavities coupled to transmon ancilla have demonstrated long intrinsic lifetimes~\cite{reagor_quantum_2016}, universal state control~\cite{heeres_implementing_2017}, and, importantly, have a single dominant error mechanism --- single photon loss --- that has been successfully mitigated through the use of quantum error correction codes~\cite{ofek_extending_2016}.  Furthermore, there are many available bosonic codes~\cite{gottesman_encoding_2001, mirrahimi_dynamically_2014, michael_new-class_2016, albert_performance_2018}, such that the choice of encoding can be optimized for specific applications and local error models.  However, up to now, this required developing tailored entangling operations for each encoding~\cite{rosenblum_a-cnot_2018,chou_deterministic_2018}.  A deterministic, codeword-independent entangling gate would be a powerful tool for connecting quantum memories that each harness the strength and flexibility of bosonic encodings in 3D cQED.

A generalized two-mode entangling operation can be enacted by performing a quantum superposition of the identity and SWAP gates.  A quantum Fredkin gate (or controlled-SWAP operation) enacts such a superposition controlled on the state of a third quantum mode~\cite{milburn_quantum_1989}, however it inevitably results in tripartite entanglement of the target modes with the control mode.  Instead, the exponential-SWAP (eSWAP) operation $\mathbf{U}_{\mathrm{E}}\left(\theta_c\right) = \exp\left(i\theta_c \, \mathbf{ SWAP}\right) = \left( \cos \left( \theta_c \right) \, \mathbf{I} + i \sin \left(\theta_c\right) \, \mathbf{ SWAP}\right)$, where the $\mathbf{SWAP}$ gate fully exchanges the states of the two cavities and $\theta_c$ is the ancilla rotation angle that provides full tunability of the operation, directly enacts the superposition of gates while leaving the ancillary mode unentangled~\cite{filip_overlap_2002, lau_universal_2016}.  This operation is analogous to the exchange operation between two spins, where an interaction splits the symmetric and antisymmetric eigenstates and result in a dynamical exchange of the spin states in time~\cite{petta_coherent_2005,anderlini_controlled_2007,kaufman_entangling_2015}.  The exchange operation provides an entangling gate for spins at particular interaction times and is the cornerstone of many schemes for universal quantum computation~\cite{divincenzo_universal_2000,kempe_theory_2001}.  Similarly, at $\theta_c=\frac{\pi}{4}$ the eSWAP operation is akin to $\sqrt{i\mathbf{SWAP}}$ and deterministically entangles two bosonic modes, regardless of their encoding. This powerful feature allows quantum information processing with different bosonic codewords on the same hardware, making the eSWAP operation a valuable building block for universal quantum computation with bosonic modes~\cite{lau_universal_2016}. 

Here, we devise an efficient circuit to implement the eSWAP operation in the 3D cQED architecture and showcase the first direct realization of the operation between two superconducting cavity modes controlled by a transmon ancilla.  The original eSWAP protocol utilizes two quantum Fredkin gates~\cite{lau_universal_2016}, which are severely limited by ancilla decoherence.  We implement a deterministic Fredkin gate in the same system, highlighting its vulnerability to the transmon decoherence, which is reduced in our optimized eSWAP protocol.  Using this operation, we demonstrate deterministic entanglement of the two cavity modes using Fock- and coherent-state encodings with a fidelity $\mathcal{F}\sim0.75$, without correcting for any state preparation and measurement (SPAM) errors.  We then show full control over the unitary operation by varying the parameter $\theta_c$. In particular, we highlight its action at the three primary settings of $\theta_c = \{0,\frac{\pi}{4},\frac{\pi}{2}\}$, corresponding to the identity, entangling, and full-SWAP operations, respectively. Finally, we perform full quantum process tomography for these three operations, extracting a lower bound on the process fidelity of $\mathcal{F}\sim0.85$ for the $\{0,1\}$ Fock encoding and $\mathcal{F}\sim0.60$ for the level-1 binomial encoding~\cite{supplement}.  Our results demonstrate the versatility of these operations as generalized entangling gates for universal quantum computation with bosonic modes.

Our system is designed to implement the eSWAP operation between states stored in two long-lived ($\sim$ms) bosonic quantum memories. In our system, two superconducting microwave cavities, Alice (orange) and Bob (blue), are dispersively coupled to a total of three transmons, as shown in Fig.~\ref{circuits}A. The two transmons qA and qB, each coupled to a single cavity, are used for universal control and state-tomography of individual cavities~\cite{heeres_implementing_2017, sun_tracking_2014}. A third, Y-shaped transmon, qC, dispersively couples to both Alice and Bob~\cite{wang_schrodinger_2016}. The non-linearity of its single Josephson junction enables four-wave-mixing, which is used to enact a frequency-converting bilinear coupling of the form $\mathbf{H}_{\mathrm{BS}}\propto g\,\mathbf{a} \mathbf{b}^\dagger + g^*\,\mathbf{a}^\dagger \mathbf{b}$ in the presence of two microwave drives. Recent work has shown that this engineered coupling can be used to realize a robust beamsplitter and \textbf{SWAP} operation between two stationary bosonic modes~\cite{gao_programmable_2018}.

Using this driven coupling, we realize a deterministic quantum Fredkin gate~\cite{milburn_quantum_1989}, where the states of Alice and Bob are swapped conditioned on the the state of qB.  Such a controlled-SWAP operation has only recently been demonstrated (non-deterministically) in linear optics experiments~\cite{patel_a-quantum_2016,ono_implementation_2017}.  In our system, $\mathbf{H}_{\mathrm{BS}}$ requires the satisfaction of the frequency-matching condition $|\omega_b - \omega_a| = |\omega_2 - \omega_1|$, where $\omega_{a}$ and $\omega_{b}$ are the frequencies of the cavities and $\omega_{1}$ and $\omega_{2}$ are those of the drives.  When either Alice or Bob is dispersively coupled to a transmon ancilla with $|\chi| > |g|$, this process is intrinsically dependent on the state of the ancilla.  We exploit this feature to perform a Fredkin gate on a selected set of initial states in the $\{0,1\}$ Fock encoding, as shown in the supplementary material~\cite{supplement}. We estimate the quality of the operation to be $\ge0.68$, uncorrected for state preparation and measurement (SPAM) errors.  A major source of imperfection is the transmon decoherence ($T_{2}\sim 30\,\mu$s) because qB must remain in a coherent superposition during the entirety of the operation.  We emphasize that implementing an eSWAP using two Fredkin gates and an ancillary control rotation, as proposed in~\cite{lau_universal_2016}, would impose a strict limit to its performance. 

Our implementation of the eSWAP operation ameliorates this penalty by minimizing both the total gate time and the duration over which the transmon is in a superposition. This is achieved by first decomposing the Fredkin  gate into two BS and a controlled phase gate (CPS) as shown in Fig.~\ref{circuits}B. Because $\mathbf{U_{\mathrm{BS}}^{\dagger}}\mathbf{U_{\mathrm{BS}}} = \mathbf{I}$, we eliminate two of the BS operations and reduce the total gate time~\cite{supplement}. Additionally, we may commute the remaining two BS operations with the transmon rotations such that the final implementation, shown in Fig.~\ref{circuits}C, keeps the ancilla in the ground state during the relatively slow BS operations. The CPS gates are realized using the dispersive coupling between qB and Bob, which imparts a $\pi$-phase to each photon in Bob when qB is excited for a period of $\sim\pi/\chi$. Therefore, we have effectively reduced the susceptibility to transmon T1 and T2 errors during the operation from $\mathcal{O}[\gamma_t(t_{\mathrm{BS}} + t_\mathrm{CPS} + t_{\mathrm{rot}})]$ to  $\mathcal{O}[\gamma_{t}(t_{\mathrm{CPS}} + t_{\mathrm{rot}})]$, where $\gamma_t$ is the transmon decoherence rate and $t_{\mathrm{BS}}\,(\sim5\,\mu$s), $t_\mathrm{CPS}\,(\sim0.5\,\mu$s), and $t_{\mathrm{rot}}\,(\sim50$\,ns) are the duration of the BS, CPS, and transmon rotations, respectively. Furthermore, this implementation has the potential to be made tolerant of these errors by using higher levels of the transmon~\cite{girvin_inprep_2018}.

 %--------Figure 2:--------------
\begin{figure}[t]
	\includegraphics[scale=1]{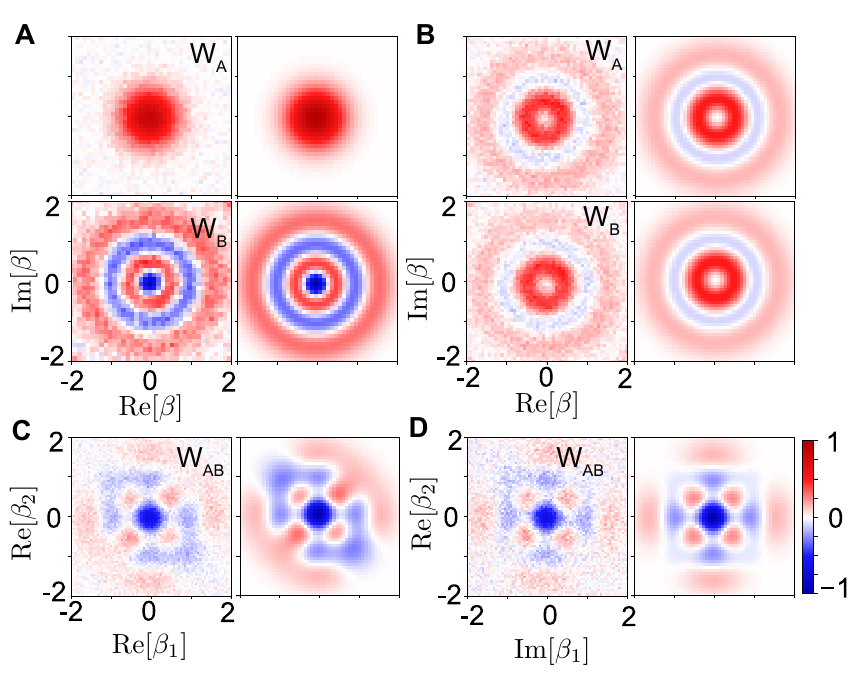}
	\caption{\textbf{Deterministic entanglement in Fock encoding.} 
		\textbf{(A)} The measured (left) and ideal (right) single Wigner functions of Alice ($W_{\mathrm{A}}$) and Bob ($W_{\mathrm{B}}$) after preparing $\ket{0}_{\mathrm{A}}\otimes\ket{3}_{\mathrm{B}}$. 
		\textbf{(B)} Left: the measured $W_{\mathrm{A}}$ and $W_{\mathrm{B}}$ after $\mathbf{U}_{\mathrm{E}}\left(\theta_c=\pi/4\right)$. right: the calculated ideal Wigners for a statistical mixture of $\ket{0}$ and $\ket{3}$.
		\textbf{(C, D)} Left: The measured joint Wigner function ($W_{\mathrm{AB}}$) after the operation $\mathbf{U}_{\mathrm{E}}\left(\theta_c=\pi/4\right)$ on Re-Re and Im-Im planes. Right: the ideal $W_{\mathrm{AB}}$ on the same planes for the maximally entangled state $\frac{1}{\sqrt{2}} \left(\ket{0}_A\otimes\ket{3}_B + i \ket{3}_A\otimes\ket{0}_B\right)$. Negativity at the origin indicates odd joint parity, which is preserved by the operation.}
	\label{fig:tomographyDemo}
\end{figure}
%----------------------------------

The most useful feature of the eSWAP operation is the ability to entangle two bosonic modes regardless of their encoding. We demonstrate this by enacting the operation $\mathbf{U}_{\mathrm{E}}(\theta_{c}=\frac{\pi}{4})$ on the input state $\Psi_\text{in} = \ket{0}_A\otimes\ket{3}_B$. We then perform Wigner tomography on Alice and Bob individually~\cite{sun_tracking_2014} to characterize their states before and after the operation (Fig.~\ref{fig:tomographyDemo}A and \ref{fig:tomographyDemo}B). We observe that the initial (separable) state in each cavity has well-defined individual parity with $\langle P_{A}\rangle \approx + 0.94$ and $\langle P_{B}\rangle \approx - 0.9$. This indicates an odd joint parity with $\langle P_{AB}\rangle \approx -0.85$. We expect the final state after the operation to be a two-mode maximally entangled state $\Psi_\text{out} = \frac{1}{\sqrt{2}} \left(\ket{0}_A\otimes\ket{3}_B + i \ket{3}_A\otimes\ket{0}_B\right)$. The measured single-cavity Wigner functions suggest that the final states do not have a well-defined individual parity. This is an indication that after the operation, the joint cavity state is no longer separable and the independent measurements of each mode erases the coherence between them. Indeed, the measured $W_{\mathrm{A}}$ and $W_{\mathrm{B}}$ show good agreement with the ideal Wigner functions of a statistical mixture of $|0\rangle$ and $|3\rangle$. 

In order to charaterize the two-mode entangled state after the operation, we must consider the joint Wigner function $W_{\mathrm{AB}}$ of Alice and Bob. We extract $W_{\mathrm{AB}}$ via shot-to-shot correlation of the displaced photon number parity measurements~\cite{wang_schrodinger_2016, supplement}. We present $W_{\mathrm{AB}}$ on the $\mathrm{Re}[\beta_1]\text{-}\mathrm{Re}[\beta_2]$ and $\mathrm{Re}[\beta_1]\text{-}\mathrm{Im}[\beta_1]$ planes in Fig.~\ref{fig:tomographyDemo}C and D, respectively. Our data (left) show excellent agreement with the ideal (right) for $\Psi_\text{out}$. The strong negativity at the origin indicates the preservation of negative joint parity by the entangling operation. The measured joint parity, $\langle P_{AB}\rangle \approx -0.75$ (uncorrected for SPAM errors) also provides a lower bound on the quality of the entangling operation~\cite{supplement}.  

%--------Figure 3:--------------
\begin{figure*}[!t]
	\floatbox[{\capbeside\thisfloatsetup{capbesideposition={right,center},capbesidewidth=5.4cm}}]{figure}[\FBwidth]
	{\includegraphics[scale=1]{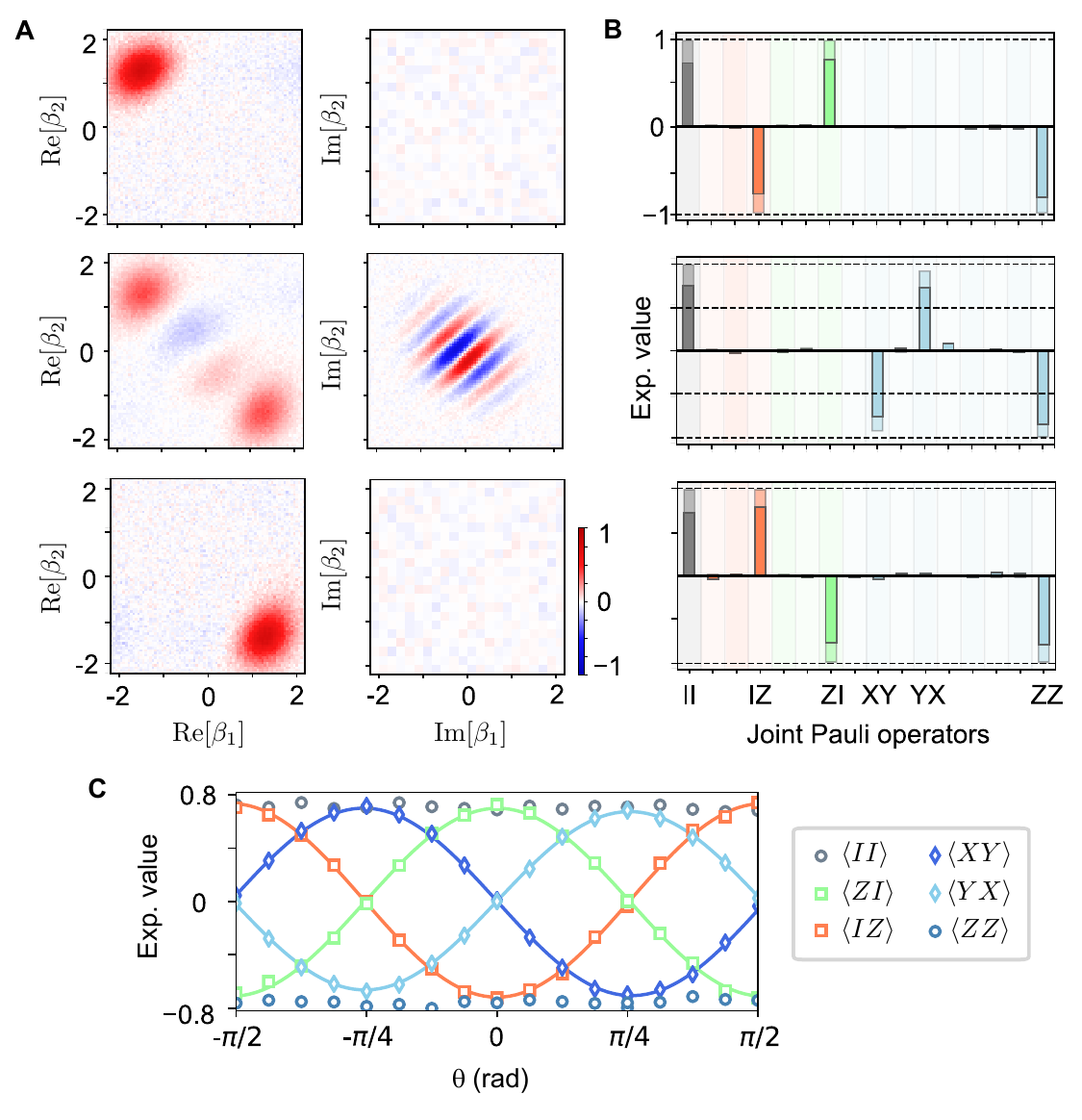}}
	{\caption{\textbf{Characterisation of eSWAP unitary in coherent state encoding.} 
		\textbf{(A)} The joint Wigner measurements in the Re-Re, Im-Im planes after $\mathbf{U}_{\mathrm{E}}(\theta=0,\pi/4,\pi/2)$, respectively. 
		\textbf{(B)} The measured (solid bars) two-qubit Pauli operators for each of the process. They show good agreement with the expected outcomes (transparent bars) for $\alpha\approx 1.41$. The ideal contains non-unity values for  $\langle XY \rangle$ and $\langle YX \rangle$ due to non-orthogonality of the basis states at this chosen $\alpha$.  
		\textbf{(C)} The expectation values of selected Pauli operators as we continuously vary the control angle, $\theta_c$.}
	\label{fig:coherentStates}}
\end{figure*}
%----------------------------------

Next, we highlight the codeword-independence of the eSWAP operation by investigating its action on Alice and Bob in the coherent state basis. To do so, we prepare the cavities in $\Psi_\text{in} \propto |\text{-}\alpha\rangle_A\otimes|\alpha\rangle_B$ (with $\alpha=1.41$) and perform the eSWAP operation. In particular, we choose to focus on three primary instances, namely, the identity ($\theta_c = 0$), entangling ($\theta_c = \frac{\pi}{4}$), and full \textbf{SWAP} ($\theta_c = \frac{\pi}{2})$ operations. Joint Wigner functions measured on the $\mathrm{Re}\text{-}\mathrm{Re}$ and $\mathrm{Im}\text{-}\mathrm{Im}$ planes for each of the final states are shown in Fig.~\ref{fig:coherentStates}A. The $\mathrm{Re}\text{-}\mathrm{Re}$ plane indicates the population distribution while the $\mathrm{Im}\text{-}\mathrm{Im}$ highlights the quantum coherence between the two modes. The fringes are only present in the $\theta_c = \pi/4$ case, where the output state is a parity-less two-mode entangled cat state: $\Psi_\text{AB} \propto |\text{-}\alpha\rangle_A|\alpha\rangle_B + i|\alpha\rangle_A|\text{-}\alpha\rangle_B$~\cite{wang_schrodinger_2016}. The features between the two population components in the $\mathrm{Re}\text{-}\mathrm{Re}$ plane can be accounted for by the self-Kerr non-linearities of Alice and Bob~\cite{supplement}. With $\theta_c=0$, we observe that the population remains in $|\text{-}\alpha\rangle_A\otimes|\alpha\rangle_B$ and at $\theta_c=\pi/2$, it is fully transferred to $|\alpha\rangle_A\otimes|\text{-}\alpha\rangle_B$. 

The state of Alice and Bob can be considered as a pair of continuous-variable qubits encoded in the coherent state basis and can be fully characterized by measuring their two-qubit Pauli operators (correlators). Using the technique described in Ref.~\cite{wang_schrodinger_2016}, we do so efficiently by probing $\langle P_{\mathrm{AB}}\rangle$ at 16 selected points of the phase space. The encoded two-qubit tomography after identity, entangling, and \textbf{SWAP} are shown in Fig.~\ref{fig:coherentStates}B. For the case of identity and \textbf{SWAP}, we find exclusively single-qubit Pauli operators. In contrast, for the entangling case, only two-qubit operators are present, indicating strong non-classical correlations between the two modes. Based on this, we obtain a direct fidelity estimation~\cite{flammia_direct_2011} of $\frac{1}{4} (\langle II \rangle - \langle XY \rangle + \langle YX \rangle - \langle ZZ \rangle) \approx 74\,\%$ to the ideal entangled state $\Psi_\text{AB} $.  

Next, we demonstrate the full tunability of the eSWAP operation by probing a selected set of two-qubit Pauli operators as a function of $\theta_c$ (Fig.~\ref{fig:coherentStates}C).  The measurement outcomes corresponding to $\langle II \rangle$, $\langle ZZ \rangle$ show no dependence of $\theta_c$, indicating the preservation of photon numbers at all angles. Oscillations of $\langle IZ \rangle$,  $\langle ZI \rangle$ are anti-correlated, consistent with the transfer of population between the modes. Finally, $\langle XY \rangle$ and $\langle YX \rangle$, which indicate two-qubit correlations, are exactly $\pi/4$ out of phase with $\langle IZ \rangle$($\langle ZI \rangle$) and show maximum contrast at $\theta_c = \pm\,\pi/4$ where the single-qubit operators vanish. This is in good agreement with the expected signature of eSWAP and showcases our ability to tune the operation via a single ancilla rotation, in analogy to varying the interaction time in an exchange gate.

%%--------Figure 4:--------------
\begin{figure}[!t]
	{\includegraphics[scale=1]{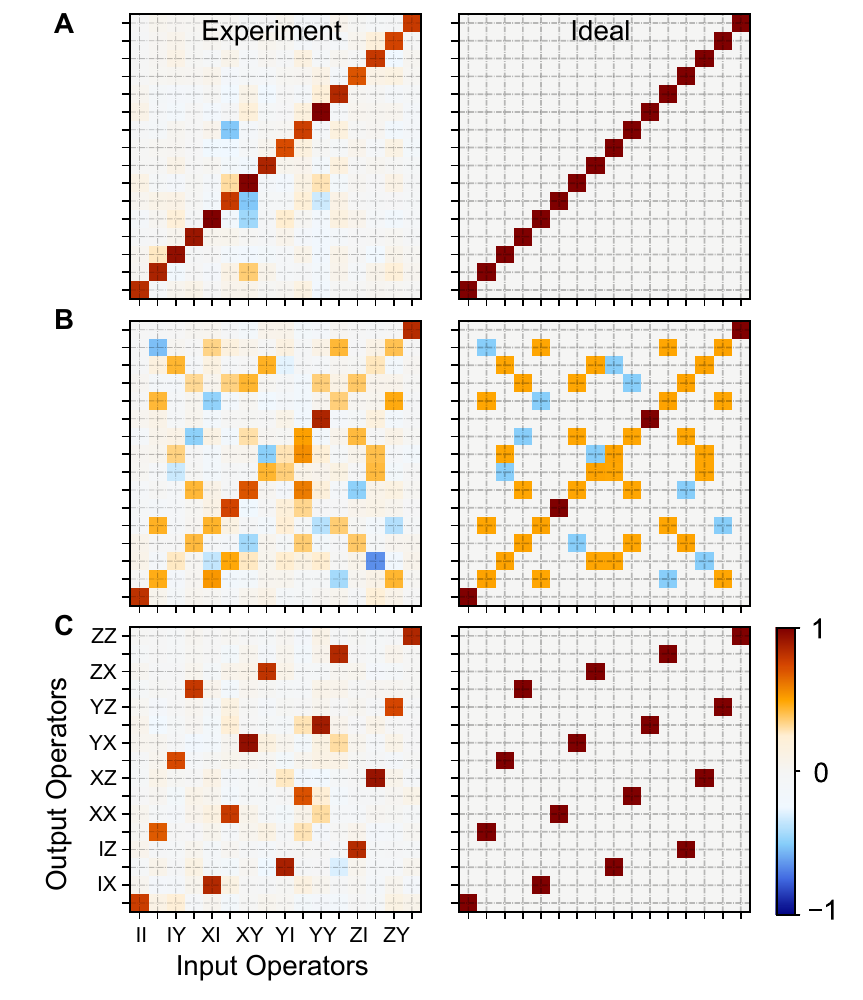}}
	{\caption{\textbf{Quantum process tomography} 
		\textbf{(A)}The experimentally reconstructed (left) and ideal (right) quantum process tomography of $\mathbf{U}_{\mathrm{E}}(0)$ in the $\{0,1\}$ Fock encoding. We represent the quantum process in the Pauli transfer representation (Supplementary Information). With all SPAM errors included, we obtain $\mathcal{F}_{0} \approx 86\%$. 
		\textbf{(B)} The reconstructed (left) and ideal (right) quantum process tomography of $\mathbf{U}_{\mathrm{E}}(\frac{\pi}{4})$ with $\mathcal{F}_{\frac{\pi}{4}} \approx 84\%$. 
		\textbf{(C)} The reconstructed (left) and ideal (right) quantum process tomography of $\mathbf{U}_{\mathrm{E}}(\frac{\pi}{2})$ with $\mathcal{F}_{\frac{\pi}{2}} \approx 83\%$. 
		} 
	\label{fig:qpt}}
\end{figure}
%----------------------------------

Finally, we perform full quantum process tomography (QPT) to obtain a quantitative analysis of our engineered operation. This is accomplished by applying the operation to sixteen input states that together span the chosen code space. We extract the density matrices of the resulting states from their joint Wigner functions and reconstruct the process matrices in the Pauli transfer representation $\mathcal{R}_{\mathrm{E}}(\theta_c)$~\cite{chow_universal_2012}. It captures the action of the operation on any given set of input and output Pauli vectors, $P_{\mathrm{in, out}}$: $P_{\mathrm{out}} = \mathcal{R}_{\mathrm{E}}(\theta_c)P_{\mathrm{in}}$.  The measured and ideal $\mathcal{R}_{\mathrm{E}}(\theta_c)$ for Alice and Bob in the $\{0, 1\}$ Fock encoding are shown in Fig.~\ref{fig:qpt} for the angles $\theta_c = \{0,\pi/4, \pi/2\}$, respectively. Our results show good qualitative agreement with the expected processes.  From this, we can calculate a process fidelity of $\mathcal{F}_{\mathrm{E}} = (84\pm2)\,\%$ averaged over the three control angles, without correcting for SPAM errors. To estimate the non-idealities due to imperfect state preparation and measurement, we perform the same procedure for the process consisting of only the encoding and measurement. This yields a process fidelity of $\mathcal{F}_{\mathrm{encode}} = (88\pm2)\,\%$, suggesting that our measured $\mathcal{F}_{\mathrm{E}}$ is likely limited by SPAM errors for this encoding. 

A crucial advantage of the eSWAP operation is its compatibility with error-correctable multi-photon encodings. To verify this capability, we perform QPT for the same three gates with both Alice and Bob encoded in the binomial basis~\cite{michael_new-class_2016}. The resulting process matrices again show good qualitative agreement with the expectation and obtain an average $\mathcal{F}_{\mathrm{E}} \approx 60\%$ with a $\mathcal{F}_{\mathrm{encode}} = (77\pm2)\,\%$~\cite{supplement}. This indicates that our current implementation of eSWAP is more susceptible to errors when large photon number states are present in the cavity modes. We attribute this to three potential sources, namely, enhanced photon loss rate during the parametrically-driven beamsplitter operations, self-Kerr non-linearities of each mode, and the magnified susceptibility to small imperfections in the CPS gates. In general, the primary limitations arise from the strong parametric drives used to engineer the beamsplitter. It is possible to mitigate the associated imperfections by developing more sophisticated mixing elements. A detailed error budget is presented in supplementary material~\cite{supplement}.

In this report, we present an efficient circuit for the eSWAP operation that can be adapted to any harmonic oscillator degrees of freedom coupled to non-linear aniclla. We demonstrate the first experimental realization of the eSWAP between two bosonic modes in cQED and showcase the deterministic entanglement of the state of two cavities encoded in Fock- and coherent-state bases. Together with single-mode gates, this provides a universal gate set on error-correctable qubits encoded in multi-photon states of cavities. Moreover, we may increase the number of cavities to implement the four-mode eSWAP gate, which will enable quantum information processing with different bosonic encodings on the same hardware~\cite{lau_universal_2016}. Our results also highlight the codeword-independent nature of the eSWAP operation. This enables us to exploit the strength of various bosonic encoding schemes within the same system and optimize the complexity for different local error models. The eSWAP operation provides both a key primitive for universal quantum computation using bosonic modes in cQED, as well as a powerful tool for the future implementation of quantum principal component analysis~\cite{lloyd_quantum_2014} and quantum machine learning~\cite{lau_quantum_2017}.

\acknowledgments{
We thank Radim Filip for helpful discussions; N. Frattini, K. Sliwa, M.J. Hatridge, and A. Narla for providing the Josephson Parametric Converters (JPCs); N. Ofek and P. Reinhold for providing the logic and control interface for the field programmable gate array (FPGA) used in of this experiment. This research was supported by the U.S. Army Research Office (W911NF-14-1-0011). Y.Y.G. was supported by an A*STAR NSS Fellowship; B.J.L. is supported by Yale QIMP Fellowship; S.M.G. by the National Science Foundation (DMR-1609326); L.J. by the Alfred P. Sloan Foundation (BR 2013-049) and the Packard Foundation (2013-39273). Facilities use was supported by the Yale Institute for Nanoscience and Quantum Engineering (YINQE), the Yale SEAS cleanroom, and the National Science Foundation (MRSECDMR-1119826).
}

%%%%%%%%%%%%%%%%%%%%%%%%%%%%%%%%%%%%%%%%%%%%%%%%%
%%%%%%%%%%%%%%%%%%%%%%%%%%%%%%%%%%%%%%%%%%%%%%%%%
%%%%%%%%%%%%%%%%%%%%%%%%%%%%%%%%%%%%%%%%%%%%%%%%%

%

%%%%%%%%%%%%%%%%%%%%%%%%%%%%%%%%%%%%%%%%%%%%%%%%%
%%%%%%%%%%%%%%%%%%%%%%%%%%%%%%%%%%%%%%%%%%%%%%%%%
%%%%%%%%%%%%%%%%%%%%%%%%%%%%%%%%%%%%%%%%%%%%%%%%%

\clearpage

\onecolumngrid
\begin{center}
\textbf{\large Supplementary Materials for: \\ 
		``Measurement-based entanglement of noninteracting bosonic atoms''
		}

\end{center}
\twocolumngrid

%%%%%%%%%%%%%%%%%%%%%%%%%%%%%%%%%%%%%%%%%%%%%%%%%
%%%%%%%%%%%%%%%%%%%%%%%%%%%%%%%%%%%%%%%%%%%%%%%%%
%%%%%%%%%%%%%%%%%%%%%%%%%%%%%%%%%%%%%%%%%%%%%%%%%
\setcounter{equation}{0}
\setcounter{figure}{0}
\setcounter{table}{0}
\setcounter{page}{1}
\makeatletter
\renewcommand{\theequation}{S\arabic{equation}}
\renewcommand{\thefigure}{S\arabic{figure}}
\renewcommand{\thetable}{S\Roman{table}}

%======================================================
%======================================================
%======================================================
%======================================================

\section{Supplementary Materials}
\subsection{Device architecture and system parameters}

%%--------Figure S1:--------------
\begin{figure}[!b]
\centering
\includegraphics[scale=0.7]{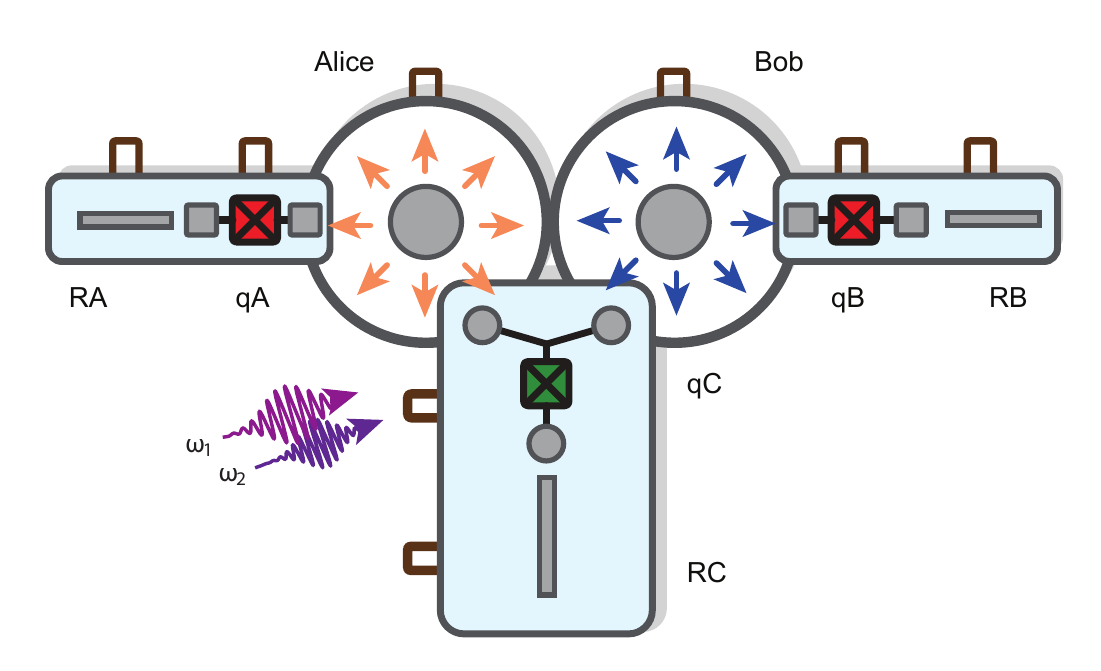}
\caption{\textbf{A cartoon showing the top view of the 3D double-cavity cQED system}. The center transmon ancilla provides nonlinear coupling between Alice and Bob. The package accommodates two additional transmon ancilla, qA and qB, which are each coupled to one of the cavities Alice and Bob, respectively.  The RF drives are coupled to the system through the drive port of qC.}
\label{sfig:cartoon}
\end{figure}
%----------------------------------

Our cQED system includes two three dimensional (3D) superconducting microwave cavities, Alice and Bob, three transmon-type devices, and three quasi-planar readout resonantors.  All components are housed in a  single block of high-purity (4N) aluminum in the structure shown in Fig.~\ref{sfig:cartoon}, which is chemically etched after machining to improve the surface quality. Alice and Bob act as quantum memories that are capable of coherently storing quantum information in bosonic states. They are formed by 3D coaxial transmission lines that are short-circuited on one end and open-circuited on the other by virtue of a narrow circular waveguide \cite{reagor_quantum_2016}. The resonance frequency of the cavities' fundamental modes are determined by the lengths of the transmission lines (4.8 and 5.6 mm respectively for Alice and Bob).  

An elliptical tunnel is machined between Alice and Bob, allowing the insertion of a chip containing the a Y-shaped transmon ancilla, qC, and its readout resonator into the cavities. Two additional tunnels are machined on either side of Alice and Bob to allow the incorporation of additional transmons, qA and qB, together with their respective readout channels.  The superconducting transmons are fabricated on sapphire substrates using electron-beam lithography and a standard shadow-mask evaporation of Al/AlOx/Al Josephson junction. During the same fabrication process, a separate strip of the tri-layer film is also deposited. Together with the wall of the tunnel, it forms a hybrid planar-3D $\lambda$/2 stripline resonator that is capacitively coupled to the transmon.  This design combines the advantages of both precise, lithographic control of the critical dimensions and the low surface/radiation loss of 3D structures \cite{axline_an-architecture_2016}. The chip containing these structures is inserted into the tunnel such that the transmon antenna(s) protrudes into the cavities to the desired capacitive coupling to Alice and Bob. Each mode is coupled to the fridge input/output lines via standard SMA couplers. 

This system is an extension of the devices used in Ref.~\cite{wang_schrodinger_2016, gao_programmable_2018}. The Y-shaped transmon is used solely as the mixing element to enact the frequency-converting bilinear coupling between Alice and Bob~\cite{gao_programmable_2018}. A single ancilla with an independent readout resonator, is coupled to each cavity in order to provide fast, independent cavity manipulations and tomography. Additionally, qB serves as the ancillary mode that controls the eSWAP operation angle. The parameters of all relevant components are summarized in Table~\ref{table:params_multi_ancilla}. We also characterize the coherence of each component in the system using standard cQED measurements. The results are summarized in Table~\ref{table:t1t2s}. 

%======================================================
%------------------------------------
%=== Table of parameters (S1):
%------------------------------------
\begin{table*}[!htb]
\centering  
%\vspace{2ex}
  	\begin{tabular}{*{7}{c}}
	\hline\hline\\[-2ex]
		\enskip Element  \enskip & \enskip Frequency	\enskip&  \multicolumn{5}{ c}{\enskip Nonlinearity to element: $\chi_{ij}/2\pi$ (MHz)  \enskip}   \\ 
		\cline{3-7}
		 \enskip&\enskip  $\omega/2\pi$ (MHz) \enskip& \enskip Alice \enskip&\enskip Bob \enskip&\enskip qA \enskip&\enskip qB \enskip&\enskip qC \enskip \\ \hline 
		Alice       	& 5467.25  & $\lesssim 0.005$ & --- & 0.79 & --- & 0.37 \\ %\hline
		Bob        	& 6548.18  & --- & $\lesssim 0.005$ & --- & 1.26 & 0.30 \\ %\hline
		qA     	& 4602.56  & 0.79 & --- & 174.20 & --- & ---  \\ %\hline
		qB     	& 4944.66  & --- & 1.26 & --- & 178.34 & ---  \\ %\hline
		qC  		& 5985.56  & 0.37 & 0.30 & --- & --- & 71.25 \\ %\hline
		RA  		& 7724.58  & --- & --- & $\sim 1$ & --- & --- \\ %\hline
		RB  		& 7722.30  & --- & --- & --- & $\sim 1$ & --- \\ %\hline
		RC  		& 8062.70  & --- & --- & --- & --- & --- $\sim 1$ \\ \hline
  	\end{tabular}
	\caption{\textbf{Hamiltonian parameters of all cQED components.} Values that are within a parenthesis are estimated/simulated parameters. Some non-linear couplings, such as $\chi$ between qA and Bob, are omitted because they are too small to be simulated and measured.}
	\label{table:params_multi_ancilla}
\end{table*}
%======================================================

%======================================================
%------------------------------------
%=== Table of coherences: (S2)
%------------------------------------
\begin{table}[!htb]
\centering  
\vspace{2ex}
\begin{tabular}{*{5}{c}}
\hline\hline\\[-2ex]
		\enskip Element   \enskip& \enskip T1 ($\mu$s) \enskip&\enskip T2 ($\mu$s) \enskip&\enskip T2$_e$ ($\mu$s) \enskip&\enskip $P_e$ \enskip  \\ \hline 
		Alice       	& 200-300 & 350-400 & --- & $\lesssim 0.01$  \\ %\hline
		Bob        	& 300-350 & 450-500 & --- & $\lesssim 0.01$  \\ %\hline
		qA     	&  45-55 & 5-10 & 40-50 & $\lesssim 0.02$  \\ %\hline
		qB     	&  70-80 & 25-35 & 75-85 & $\lesssim 0.04$   \\ %\hline
		qC  		&  10-20 & 8-16 & 15-20 & $\lesssim 0.01$   \\ %\hline
% [1ex] adds vertical space
\hline
\end{tabular}
\caption{\textbf{Coherence properties of the the system.} The device exhibits some fluctuations in its coherence times. The relatively low $T_2$ of qA and qC are likely a result of low-frequency mechanical vibrations. The coherence properties of Alice and Bob during the time of this experiment are inferior compared to those of the same device several thermal cycles ago ($T_1\sim 1$ms). This is likely a result of the degradation of its surface quality (through oxidation and thermal expansion/contraction cycles). We believe that the quality factors can be improved by chemically treating the surface of the cavities.}\label{table:t1t2s}
\end{table}
%======================================================

\subsection{Joint Wigner tomography and state reconstruction}
We characterize the collective state of Alice and Bob using joint Wigner tomography, which is a measurement of their displaced joint photon number parity, $P_{J}(\beta_1, \beta_2)$ in a four-dimensional phase space spanned by the complex numbers $\beta_1$ and $\beta_2$. In Ref.~\cite{wang_schrodinger_2016} measurements of the joint parity (and thus the joint Wigner function, $W_{\mathrm{AB}}$) were performed using the Y-shaped transmon, qC, that dispersively couples to both Alice and Bob. This relies on using multiple-levels of qC to achieve an effective matched $\chi$ to each cavity mode, which requires a relatively stringent set of system parameters. 

In our system, the availability of the two ancilla, qA and qB, and their independent single-shot readout enables a simpler joint Wigner measurement. We simultaneously map the parity of Alice and Bob to their respective ancilla and perform joint single-shot readout on qA and qB. This yields the individual displaced parity $P_{\mathrm{A}}(\beta_1)$ and $P_{\mathrm{B}}(\beta_2)$. We then extract the joint displaced parity $P_{\mathrm{AB}}(\beta_1, \beta_2) = P_{\mathrm{A}}(\beta_1)P_{\mathrm{B}}(\beta_2)$ by multiplying the two individual measurement outcomes in each run of the experiment. From this, we obtain the four-dimensional two-mode Wigner function $W_{\mathrm{AB}} = \langle P_{\mathrm{AB}}(\beta_1, \beta_2)\rangle$, which fully characterizes the joint state of Alice and Bob up to the cutoff number of photons, $N_{\mathrm{cutoff}}$, which is chosen based on specific encoding. From this, we can reconstruct the density matrix $\rho_{\mathrm{AB}}$ in the restricted Hilbert space using standard techniques~\cite{wang_schrodinger_2016}. 

We do not constrain the trace of the extracted density matrix to be unity to avoid making any a priori assumptions about the different sources of imperfection. Failures of the state preparation, tomography, and the operation will all show up as a reduced trace and final state fidelity. For the $\{0, 1\}$ Fock state encoding, we obtain an average state fidelity across the 16 basis states of $\sim 85\%$ with the operation and $\sim 88\%$ without. This provides an estimate of the reduction in state fidelity purely due to the state preparation and measurement imperfections. Additionally, we can infer the non-idealities due to the joint Wigner tomography by considering the state $|0\rangle_{\mathrm{A}}|0\rangle_{\mathrm{B}}$. This yields a state fidelity of $\sim 90\%$, consistent with the maximum contrast of the individual parity measurements of Alice ($\sim 0.95$) and Bob ($\sim 0.95$). This is limited by the readout errors ($\sim 2\%$), ancilla decoherence ($\sim 2\%$), and imperfections in the transmon rotation pulses ($\sim 1\%$). 

Similar analysis is done for the binomial encoding, which has an average photon number $\bar{n} = 2$ in each mode. It yields an average fidelity of $\sim 80\%$ for the initial states without the eSWAP operations. This degradation is primarily due to the longer optimal control theory (OCT) pulses ($\bar{n}/\chi \sim 1\,\mu$s) required to prepare the initial states~\cite{heeres_implementing_2017}, which would fail if the transmon dephases. Therefore, this process is limited by the $T_2$ of qA and qB, which are relatively low in this particular sample compared to typical transmons. This can be improved in future implementations with better shielding, vibration isolation, and more robust package designs. 

\section{Realization of a quantum control-SWAP (Fredkin) gate}

We can perform a full SWAP operation between two bosonic modes using the parametrically-driven bilinear coupling as discussed in Ref.~\cite{gao_programmable_2018}. This type of coupling is only resonant when the drives satisfy the frequency-matching conditions of the four-wave-mixing process. Here, this is given by $\omega_2-\omega_1 = \omega_b-\omega_a$, where $\omega_{1}$ and $\omega_2$ are the frequencies of the two drives and $\omega_a$ and $\omega_b$ are the frequencies of Alice and Bob, respectively. In our setup, Bob is dispersively coupled to qB, with strength $\chi_b$. When $|\chi_b| >> |g|$, where $g$ is the bilinear coupling coefficient, the four-wave-mixing process can be tuned in and out of resonance by the the state of qB. 

\begin{figure}[!t]
	\centering
	\includegraphics[scale=1]{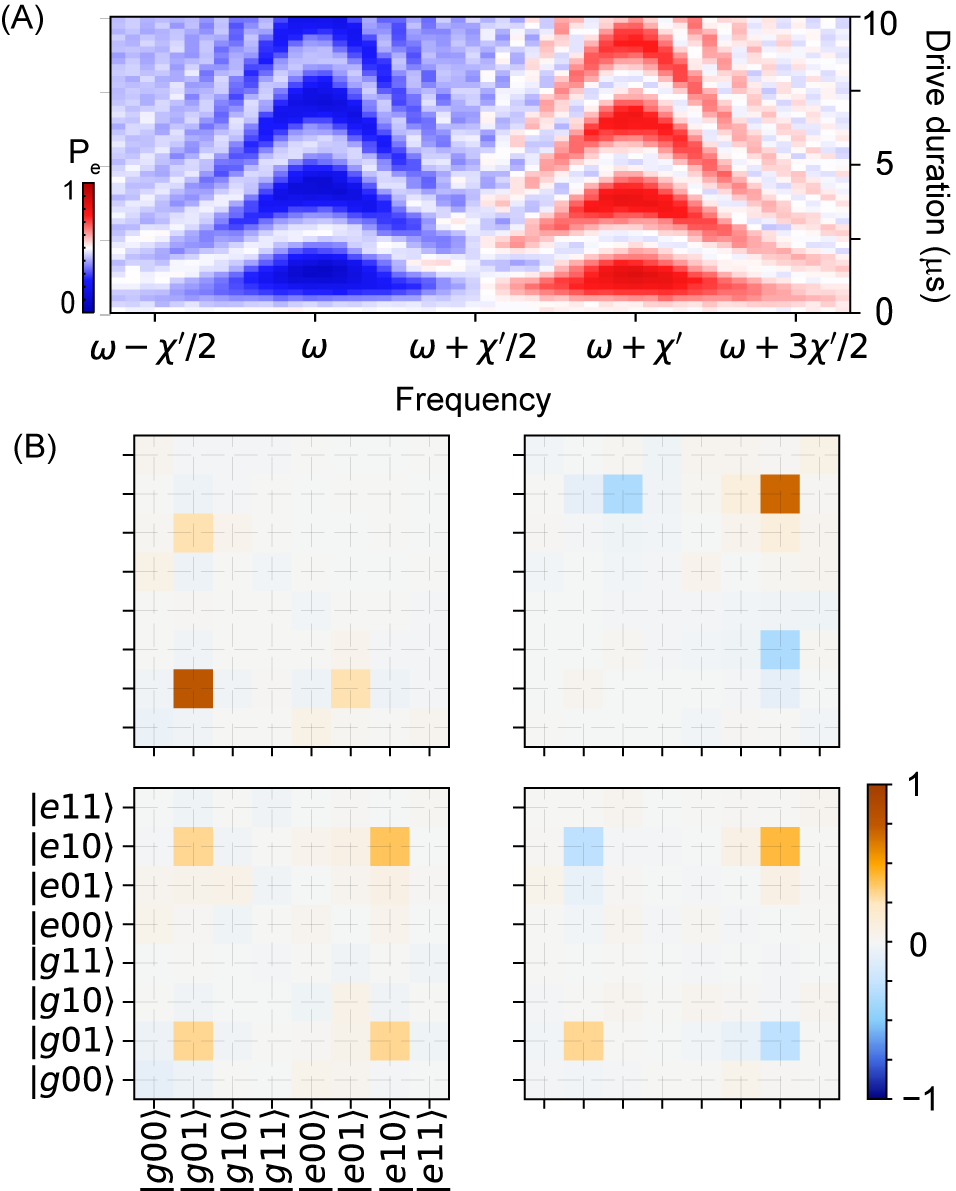}
	\caption{\textbf{A quantum Fredkin gate} 
		\textbf{(A)} The resonance condition for a parametrically driven SWAP operation measured as a function of one of the drive frequencies and the drive duration with the ancillary transmon initialized in $(|g\rangle + |e\rangle)/\sqrt{2}$. 
		\textbf{(B)} The 	reconstructed three-mode density matrix after the Fredkin gate for the initial states $|g\rangle|0\rangle|1\rangle$ (upper left), $|e\rangle|0\rangle|1\rangle$ (upper right), $\frac{1}{\sqrt{2}}(|g\rangle + |e\rangle)|0\rangle|1\rangle$ (bottom left), and $\frac{1}{\sqrt{2}}(|g\rangle - |e\rangle)|0\rangle1\rangle$ (bottm right) respectively. 
		}
	\label{sfig:cswap}
\end{figure}

To verify that this condition is sufficiently satisfied, we perform a spectroscopy experiment with qB initialized in $|+\rangle = (|g\rangle + |e\rangle )/\sqrt{2}$ and the cavities in $|0\rangle_{\mathrm{A}} |1\rangle_{\mathrm{B}}$. We then selectively excited qB conditioned on Bob being in vacuum as a function the frequency and duration of one of the two drives with the that of the other fixed as a chosen set of drive amplitudes. The results of this calibration experiment is shown in Fig.~\ref{sfig:cswap}(A), where the colorbar corresponds to the probability of qB being measured in $\ket{e}$ after the final transmon rotation pulse. This provides an indication for the successful exchange of the single excitation between Alice and Bob, and hence, acts as a meter for whether the resonance condition for the bilinear coupling is satisfied. We observe that the SWAP operation is enacted at the frequency $\omega$ and $\omega + \chi'$, which are well-separated compared to the spectral width of the resonance. We exploit this feature to realize the controlled-SWAP operation between Alice and Bob, using qB as the control mode. More specifically, we choose position the drive frequency at $\omega$, such that the bilinear coupling is resonant when qB is in $|e\rangle$. The resulting operation is described by:
\begin{equation}
\mathbf{cSWAP} = |g\rangle\langle g| \otimes \mathbf{I} + |e\rangle\langle e| \otimes \mathbf{SWAP}
\end{equation}

%%--------Figure S2:--------------
%----------------------------------

This a three-mode controlled-SWAP operation, also known as the quantum Fredkin gate~\cite{milburn_quantum_1989}. Other protocols for the implementation of such a operation have been proposed in the context of quantum optics~\cite{gong_methods_2008, fiurasek_universal_2002}. In contrast to the probabilistic nature of these protocols, our realization of the quantum Fredkin gate is fully deterministic. It produces a three-mode entangled state between Alice, Bob, and qB when qB is initialized in a superposition state $|\pm\rangle$. We characterize the action of this operation on a set of 4 initial states by implementing conditional state tomography. This is can extension from the method used in Ref.~\cite{vlastakis_characterizing_2015}. We perform a first measurement to detect the state of qB along one of its basis vectors $\{X, Y, Z\}$. Subsequently, we perform a joint Wigner measurement~\cite{wang_schrodinger_2016} of Alice and Bob to extract their joint density matrix, $\rho_{\mathrm{AB}}$. Finally, we construct the full three-mode density matrix, $\rho$, using a method described in~\cite{nielsen_quantum_2000}: 
\begin{equation}
\rho = \begin{pmatrix} \rho_1 & \rho_2\\ \rho_3 & \rho_4 \end{pmatrix}
\end{equation}
where each of the constituent part is given by the correlated transmon and cavity density matrices, with
\begin{align}
\rho_1 & = \mathcal{E}(|g\rangle\langle g|)\\
\rho_4 & = \mathcal{E}(|e\rangle\langle e|)\\
\rho_2 & = \mathcal{E}(|+\rangle\langle +|) - i \mathcal{E}(|-\rangle\langle -|) - (1-i)(\rho_1 + \rho_4)/2 \\
\rho_3 & = \mathcal{E}(|+\rangle\langle +|) + i \mathcal{E}(|-\rangle\langle -|) - (1+i)(\rho_1 + \rho_4)/2
\end{align}
where $\mathcal{E}(|k\rangle\langle k|)$ corresponds to the extracted $\rho_{\mathrm{AB}}$ after projecting into the transmon state $|k\rangle$. Using this technique, we obtain the density matrices shown in Fig.~\ref{sfig:cswap}(B) for the initial states $|g\rangle|0\rangle|1\rangle$,  $|e\rangle|0\rangle|1\rangle$,  $|+\rangle|0\rangle|1\rangle$, and $|-\rangle|0\rangle|1\rangle$ respectively. They are in good qualitative agreement with the expected density matrices, with an average overlap with $\rho_{\mathrm{ideal}}$ of $68\pm5\%$, without correcting for an SPAM errors. In particular, the bottom two plots show clear evidence of entanglement between the three modes with phases consistent with that of the initial superposition of qB. 

To the best of our knowledge, this is the first experimental realization a deterministic quantum Fredkin gate in cQED. It is a valuable tool for the implementation of universal quantum computation~\cite{lanyon_experimental_2007}, quantum cyptography~\cite{ekert_direct_2002}, and measurement~\cite{fiurasek_universal_2002}. 

\section{The simplified exponential-SWAP}

%%--------Figure S3:--------------
\begin{figure*}[!tbh]
	\centering
	\includegraphics[scale=1]{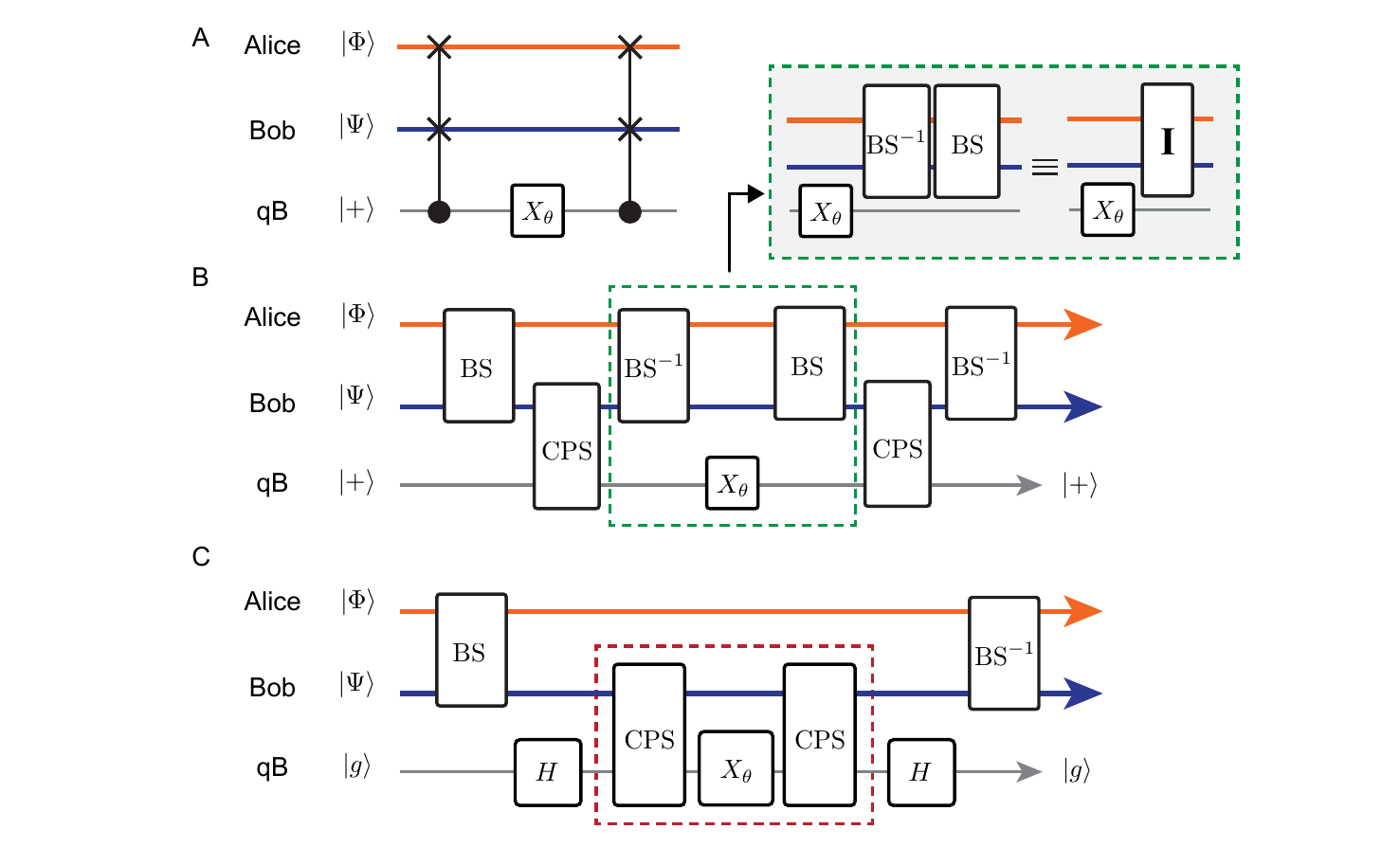}
	\caption{\textbf{The simplified eSWAP circuit} 
	\textbf{(A)} The original circuit proposed in Ref.~\cite{lau_quantum_2017} consisting of two cSWAP operations and a transmon rotation. 
	\textbf{(B)} New circuit obtained by decomposing cSWAP into two beamsplitters, $\mathbf{U}_{\mathrm{S}}$, separated by a controlled phase gate (CPS).  Inset: because the beamsplitter operation commutes with transmon rotation and $\mathbf{U}_{\mathrm{S}}^{\dagger}\mathbf{U}_{\mathrm{S}} = \mathbf{I}$, eliminate two of the beamsplitters to reduce the overall operation time. 
	\textbf{(C)} The final circuit after commuting the beamsplitter and the Hadamard gate. Now the transmon states and ends in its ground state. It is only required to remain in a coherent superposition state during the relatively fast CPS and single transmon rotation (red box). 
		}
	\label{sfig:circuit}
\end{figure*}
%----------------------------------

As proposed by Lau and collaborators in Ref.~\cite{lau_quantum_2017}, the eSWAP operation between two bosonic modes can be realized using a two controlled-SWAP gates and an ancilla rotation. A diagram of this procedure is shown in Fig.~\ref{sfig:circuit}A, where we initialize the tranmon ancilla in $|+\rangle = (|g\rangle + |e\rangle)/\sqrt{2}$ before applying two cSWAP separated by a transmon rotation around the x-axis by angle $2\theta$. At the end of this sequence, Alice and Bob undergo the eSWAP operation, with control angle $\theta$.  while the transmon ancilla is disentangled from the cavity modes and restored to $|+\rangle$. In this protocol, it is crucial that the transmon remains in a coherent superposition through the entirety of the operation. 

We have demonstrated in the previous section that although it is straightforward to realize the cSWAP operation in our system, its the performance is severely limited by the transmon decoherence. This is because a cSWAP requires the transmon to remain in a coherent superposition throughout the slow $\mathbf{SWAP}$ operation, which introduces significant non-idealities as a result of both $T_1$ and $T_2$ errors. While it is possible to speed up the $\mathbf{SWAP}$ operation by increasing the driven bilinear coupling strength, it has been shown that stronger drives tend to introduce additional sources imperfections that can further degrade the quality of the operation or cause other undesired transitions~\cite{gao_programmable_2018}. Therefore, repeating the cSWAP twice to perform the eSWAP would in general impose a heavy penalty in its performance. 

We can suppress the effects of decoherence on the quality of our eSWAP operation by reducing the overall gate time and, more importantly, the duration over which the transmon is not in its ground state. We do so by first decomposing the cSWAP intro two beamsplitters and a single controlled phase gate (CPS). 
This is analogous to a Mach-Zehnder interferometer with a controlled phase shifter on one arm.  
This leads us to the circuit shown in Fig.~\ref{sfig:circuit}B. It is important to notice that the beamsplitters and transmon operations commute with each other. This allows us to switch the order of $X_{\theta}$ and the beamsplitter such that we now have two sequential beamsplitters, which can be eliminated because $\mathbf{U}_{\mathrm{BS}}^{\dagger}\mathbf{U}_{\mathrm{BS}} = \mathbf{I}$. Finally, we switch the Hadamard used to prepare the tranmon in $|+\rangle$ and the beamsplitter so that the transmon can now remain in its ground state during the beamsplitter unitaries. With the above-mentioned changes, we arrive at the simplified circuit design for the eSWAP operation (Fig.~\ref{sfig:circuit}C).

\section{Extended data: the chi-matrix} 

%%--------Figure S4:--------------
	\begin{figure*}[t!]
		\centering
		\includegraphics[scale=1]{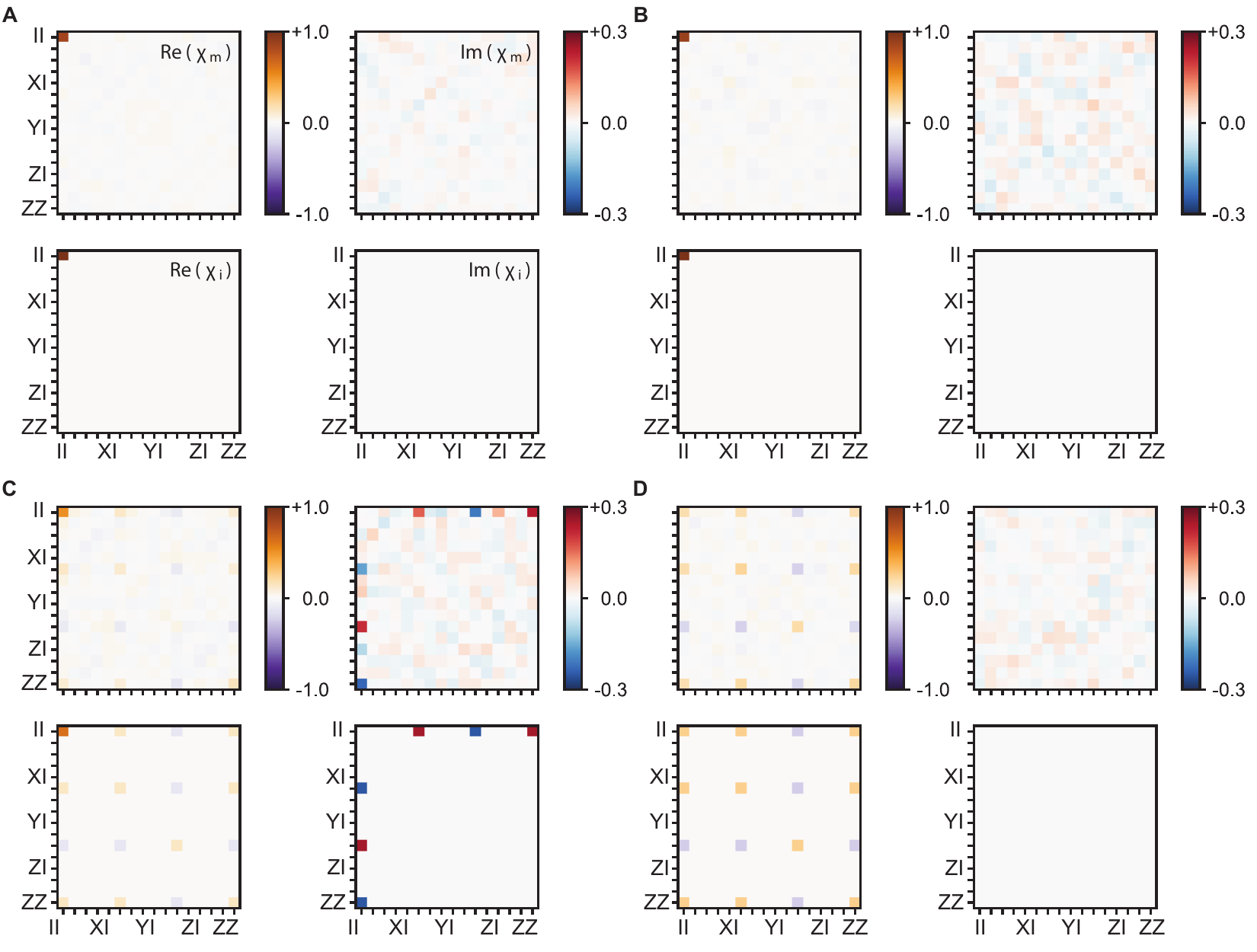}
		\caption{\textbf{Quantum process tomography.}  The real (left) and imaginary (right) components of the complex process matrix $\chi$ are shown for the (A) input, (B) identity operation, (C) $\sqrt{\mathbf{SWAP}}$ and (D) $\mathbf{SWAP}$.  For each operation, we show the measured (top) and ideal (bottom) matrices for comparison. Each of the measured process matrices are calculated without correction for SPAM errors.}
		\label{sfig:qpt_chiMatrices}
	\end{figure*}
%----------------------------------

In the main text and the analysis in the following section, we have chosen to represent the quantum process using the Pauli-transfer matrix representation $\mathcal{R}$. This is a convenient choice for the representation of a general quantum process because it is a single real-valued matrix containing complete information~\cite{chow_universal_2012}. 
However, another common representation is to reconstruct the state transfer matrix $\chi$~\cite{chuang_prescription_1996}. For general quantum processes, $\chi$ is a matrix of complex numbers, which is an over-complete representation of the process. Regardless, following the procedure described in Ref.~\cite{chuang_prescription_1996}, we have analyzed the same process data to construct $\chi$ matrices for each operation. The reconstructed $\chi$ matrices for the Fock encoding are shown in Fig.~\ref{sfig:qpt_chiMatrices}.  We obtain an average process fidelity $\mathcal{F}=\text{Tr}\left(\chi_i\chi_m\right)$ of the identity operation for state preparation (and no other operations) is $\mathcal{F}\sim 0.85$.  The process fidelities for the eSWAP operation are \textbf{0.88}, \textbf{0.82}, and \textbf{0.82}, for the operational angles $\theta=0$, $\frac{\pi}{4}$, and $\frac{\pi}{2}$, respectively. These are consistent with the analysis based on the Pauli-transfer matrices.  Because both process representations are complete, we choose to focus our analysis on the $\mathcal{R}$ representation for convenience. 

The differences in the fidelity calculated using the $\chi$ representation and the $\mathcal{R}$ representation of the process is likely due to the combination of statistical uncertainty and the difference in how errors propagate in the different representations. As one example, the identity process in the $\chi$ representation is a complex matrix with only one entry, whereas in the $\mathcal{R}$ representation it is a diagonal matrix. This difference could lead to increased sensitivity to statistical fluctuations in the case of the $\chi$ representation (by weighting only a single element very heavily and removing the effects of all other entries). 

\section{\lowercase{e}SWAP Error Budget}
\label{errorbudget}

In this work, we have demonstrated deterministic entanglement of two bosonic modes encoded in several different bases. In this section, we will discuss the primary imperfections that are present in our experiment. 
First, we discuss the state preparation and measurement (SPAM) errors that are present in each of the state and process tomography data sets presented (regardless of encoding). 

%%--------Table S3:--------------
	\begin{table*}[t]
		\centering
		\begin{tabular}{c  c c } 
		\hline\\[-2ex]
				& Assessment	&  Estimated infidelity	 \\
			\hline\hline\\[-2ex]
			transmon initialization errors & $\sim0.5\%$ probability not in $|g\rangle$	& $\sim1\%$\\
			cavity initialization errors	& $\sim0.5\%$ probability not in $|0\rangle_A|0\rangle_B$	& $\sim1\%$\\
			ancilla decoherence during OCT	& $|g\rangle$-$|e\rangle$ superposition for $t_\text{prep}^{A,B}$ & 2\% (5-8\%)\\
			parity mapping 			& population mixing in $|g\rangle$-$|e\rangle$-$|f\rangle$ rotations & $\sim1\%$	\\
					 			& finite spectral selectively			& 1\% ($\sim5$\%)	\\
			readout infidelities		& 1.0--1.5\% error per readout			& $2.5\%$\\	
			\\[-2ex] \hline\\[-2ex]
			Total	&	&	$\sim9\%$ (16-19\%)\\[0.3ex]
			
		\end{tabular}
		\caption{\textbf{Estimated infidelities due to SPAM errors.} Estimated contribution to the reduction in process fidelity for the eSWAP operation on Alice and Bob. The items and numbers in parenthesis are the additional errors when Alice and Bob are prepared in the binomial encoding. }
		\label{errorBudget_spam}
	\end{table*}
%----------------------------------

%

In our system, the initial states in Alice and Bob are prepared using numerically optimized pulses computed using the Gradient Ascent Pulse Engineering (GRAPE) method described in Ref.~\cite{heeres_implementing_2017}. This requires a simultaneous drive on each cavity and the transmon for a minimum duration of $t_{\mathrm{prep}}\sim\bar{n}/\chi$, where $\bar{n}$ is the average photon number of the desired cavity state and $\chi$ is the dispersive coupling between the cavity and its transmon ancilla. During the preparation pulse, the system is vulnerable to both the relaxation and dephasing errors of the transmon ancilla. In general, we expect an overlap of prepared initial state to the ideal of $(1-t^{A}_{\mathrm{prep}}/T_2^{A})(1- t^{B}_{\mathrm{prep}}/T_2^{B})\sim 95\%$ for an average of one photon in each cavity. 

Similarly, the joint Wigner tomography also introduces additional non-idealities due to ancilla decoherence, imperfect parity mapping, and readout errors. The estimated infidelities corresponding to each of these SPAM errors are presented in Table~\ref{errorBudget_spam}. 
This is reasonably consistent with our extracted process fidelity for the identity operation of $\sim88\%$ (Fock) and $\sim77\%$ (binomial), measured directly after the encoding pulses. In the binomial state encoding, the SPAM errors are larger, which is consistent with the expectation that both state preparation and readout fidelity are reduced with larger photon number. Primarily, we expect that the increased susceptibility to cavity dephasing and faster decay leads to reduced fidelity of the state preparation and readout of the cavity states~\cite{wang_schrodinger_2016}.

%%--------Table S4:--------------
	\begin{table*}[t]
		\centering
		\begin{tabular}{c  c l } 
		\hline\\[-2ex]
			\\[-2ex] \\[-2ex]
				& Assessment	&  Estimated infidelity	 \\
			\\[-2ex] \hline\hline \\[-2ex]
			qC excitation during operation & pump-induced heating during BS & \qquad $\lesssim 2$ \%\\			
			photon loss during operation & [$\bar{n}=0.5$ photons per cavity for 3.9 $\mu$s] & \qquad [$\sim$1 \%] \\			
								 & $\bar{n}=2$ photons per cavity for 3.9 $\mu$s & \qquad $\sim5$ \%\\			
			effective self-Kerr of cavities & estimated: $K_{\mathrm{A}} \sim 6$ kHz, $K_{\mathrm{B}} \sim 4$ kHz & \qquad $\sim8$ \% (5--15 \%) \\
			conditional phase operation &  miscalibration of phase  ($\le 2$ \%) & \qquad $\le 3$ \% \\
			ancilla excitation during operation & miscalibrated transmon rotations &\qquad $<1$ \%\\			
			\hline\\[-2ex]
			Total	&	&	\qquad13--19 \%\\[0.3ex]
			\hline
		\end{tabular}
		\caption{\textbf{Error Budget for eSWAP operation.} Estimated contribution to the reduction in process fidelity for the eSWAP operation on Alice and Bob. The  first entry is relevant for both encodings, the entry in square brackets is specific to the $\{0,1\}$ Fock encoding, and the remaining entries are specific to the Binomial encoding. }
		\label{errorBudget_operation}
	\end{table*}
%----------------------------------

Based on the above analysis, we claim that the resulting process fidelities for the eSWAP operation in the $\{0,1\}$ Fock encoding is limited primarily by SPAM errors and the operation itself only introduces an additional $\sim4\%$ infidelity. This is likely primarily from single-photon loss and spurious heating of the mixing element (qC), the first two entries in Table~\ref{errorBudget_operation}, during the two beampslitter operations.  In separate measurements, we have observed the population of qC after two BS operations to be $\lesssim 2$ \%. Additionally, single-photon-loss from the cavities (characterized ty the cavity $T_1$ from Table~\ref{table:t1t2s}) is expected to contribute $\sim1$ \% infidelity during the course of the operation. 

%%--------Figure S5:--------------
\begin{figure}[!b]
	\centering
	\includegraphics[scale=1]{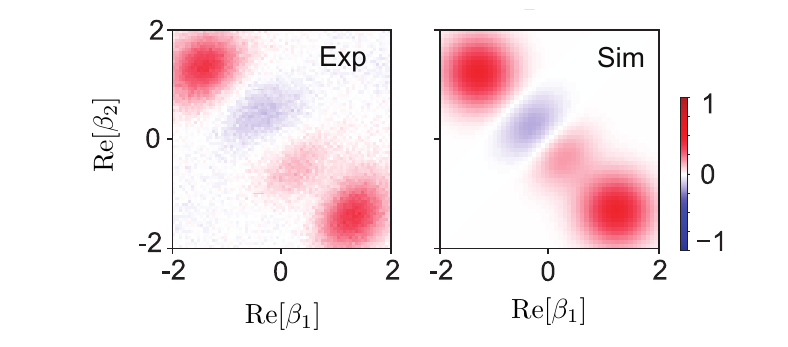}
	\caption{\textbf{Measured and simulated joint Wigner function.} Left: the measured joint Wigner function of $|\alpha\rangle|\text{-}\alpha\rangle$ after $\mathbf{U}_{\mathrm{E}}(\pi/4)$ in the Re-Re plane. Right: the corresponding simulation with $K_A/2\pi \approx 6$\,kHz and $K_B/2\pi \approx 4$\,kHz. The features in between the two population components at $|\alpha\rangle|\text{-}\alpha\rangle$ and $|\text{-}\alpha\rangle|\alpha\rangle$ reproduce that of in the experiment. }
	\label{sfig:kerr}
\end{figure}
%----------------------------------

However, for logical encodings with larger average number of photons in Alice and Bob, we expect additional non-idealities in the operation due to the presence of undesired non-linear effects (e.g., self-Kerr), decoherence of the cavity states, and enhanced susceptibility to small mis-calibrations in the operation.  In particular, the self-Kerr of the cavities results in a detuning from the resonance condition of the bilinear coupling and reduces the quality of the BS operations. The effect of this non-linearity was directly observed in the coherent state encoding with $\bar{n}\approx 2$ in each cavity, as shown in Fig.~\ref{sfig:kerr}. Here, the $W_{\mathrm{AB}}$ simulated in presence of the self-Kerr non-linearities of Alice and Bob fully reproduces the features in the experimental data. 

Furthermore, in the specific context of the Binomial encoding, we estimate the effect of the self-Kerr (using values estimated from separate coherent state measurements) on the process fidelity by simulating the operation with self-Kerr included and extracting the fidelity to the ideal state. Similarly, we estimate the effect of possible miscalibrations in the operation by including only the imperfect calibration (and no self-Kerr or other imperfection) and calculating the fidelity to the ideal state. These estimated infidelities are outlined in Table~\ref{errorBudget_operation}.
	
\begin{figure}[tb]
	\centering
	\includegraphics[scale=1]{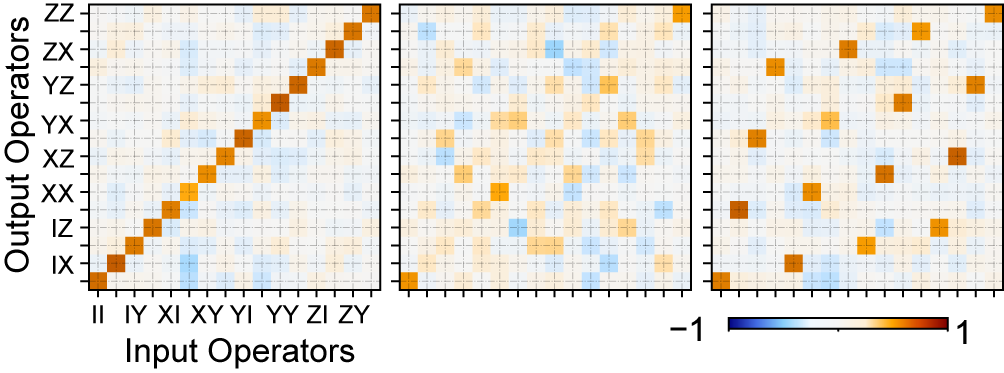}
	\caption{\textbf{Quantum process tomography.} Left to right: The process matrix in the Pauli transfer representation for $\mathbf{U}(0)$, $\mathbf{U}(\pi/4)$, and $\mathbf{U}(\pi/2)$ acting on Alice and Bob encoded in the Binomial basis. From these results, we obtain process fidelity of 0.70, 0.58, and 0.65 for the three operations without correction for SPAM errors.}
	\label{sfig:qpt_binomial}
\end{figure}

We perform quantum process tomography on three primary configuration of the eSWAP with Alice and Bob encoded in the Binomial basis using same protocol described in the main text. We show the extracted Paul transfer matrices for $\theta_c = 0, \pi/4,$ and $\pi/2$ respectively in Fig.~\ref{sfig:qpt_binomial}. They show good qualitative agreement with the ideal processes but suffers a sizable reduction in contrast. We estimate the process fidelity by performing an overlap calculation of the measured $\mathrm{R}$ with the ideal cases. This yields $\mathcal{R}_{0} \approx 0.7$, $\mathcal{R}_{\frac{\pi}{4}} \approx 0.58$, and $\mathcal{R}_{\frac{\pi}{2}} \approx 0.65$, with the a fidelity of encoding of $\mathcal{R}_{\mathrm{encode}} \approx 0.77$. 

%%--------Figure S6:--------------
%----------------------------------

Based on our previous work~\cite{gao_programmable_2018}, the strong microwave drive tones introduced to enable the tunable beamsplitters between the two stationary bosonic modes can result in both a faster photon loss rate and an stronger self-Kerr non-linearity in the cavity modes. 
From the error budget discussed in the previous section, these induced effects are expected to be the primary additional sources of error for the binomial encoding.  
The former can, in principle, be alleviated by established quantum error correction (QEC) protocols~\cite{michael_new-class_2016}, but because we do not perform QEC on the encoded states, the photon loss directly manifests as a reduction in the extracted state and process fidelity. On the other hand, the enhanced self-Kerr can not be mitigated with any currently available QEC techniques. 

One method to mitigate both of these effects is to increase the bilinear coupling strength such that the beamsplitter operations are much faster than the time scale of both the photon loss and the self-Kerr. However, stronger drives tend to introduce additional non-idealities, such as more significant cavity decoherence and spurious transitions. The details of these drive-induced effects are currently being investigated both experimentally and theoretically~\cite{zhang_inprep_2018}. Alternatively, more sophisticated mixing elements can be used to enable a faster and more robust conversion process. It has been shown that a multi-junction device such as the Superconducting Nonlinear Asymmetric Inductive eLement (SNAIL)~\cite{frattini_three_2017} or an Inductively Shunted Transmon (IST)~\cite{venkatraman_inprep_2018} can potentially improve the quality of such engineered frequency-converting processes. 

In Table~\ref{errorBudget_operation} we list the effects that reduce the fidelity of the gate operation for the binomial encoding. Despite the presence of these imperfections, our results show clear indication of the successful implementation of the eSWAP on two bosonic modes encoded in multi-photon, error-correctable codewords. The limitations of the gate fidelity are well-understood and can be suppressed via a improvements in coherence times and incorporation of more sophisticated calibration procedure.
                                                                                                                                                                                                                                                                                                                                                                                                                                                            
\end{document}